\documentclass[10pt,a4paper]{article}
\usepackage[utf8]{inputenc}
\usepackage[english]{babel}
\usepackage[T1]{fontenc}
\usepackage{amsmath}
\usepackage{amsfonts}
\usepackage{amssymb}
\usepackage{makeidx}
\usepackage{graphicx}
\usepackage{cite}
\usepackage[margin=0.9in]{geometry}
\usepackage{tikz}
\usetikzlibrary{arrows.meta}

\author{István Magashegyi$^1,$ Katalin Oltyán$^{1}$ and Péter Földi$^{1,2}$}
\date{%
    $^1$University of Szeged, Department of Theoretical Physics, Szeged, Hungary\\%
    $^2$ELI-ALPS, ELI-HU Non-profit Ltd., Szeged, Hungary\\[2ex]%
}

\title{Charges carried by wave packets \\that are superimposed on plane waves}

\begin{document}
\maketitle
\begin{abstract}
We consider the superposition of plane waves and localized wave packets. This kind of wave function can result from a local excitation of a particle described by a plane wave. For charged particles, the wave packet means a current, the time integral of which results in an extra charge, additionally to the one carried by the plane wave. When the duration of the wave packet is too short for its details to be resolved experimentally, it is the extra charge that can be determined and analyzed for information regarding the nature of the interaction that created the wave packet. Assuming that the wave function is known initially, we show an analytic method for the calculation of the extra charge by the aid of Fourier transform. Our approach is verified by comparing to both the well-known case of a Gaussian wave packet and numerically obtained results. As an important physical example, finally we consider the case of local excitation by laser pulses.
\end{abstract}

\section{Introduction}
Quantum mechanical wave functions are known to correspond to probability current densities, which, in case of charged particles, are directly related to electric currents \cite{S94}. When the charged particle is represented by a moving wave packet, one can calculate the corresponding time-dependent current that flows through a given, fixed surface. Integrating this current, the charge that can be collected along the surface is obtained. This method is relatively straightforward, although the actual calculations sometimes can only be performed numerically. Since it is the calculation of the time evolution that is numerically expensive, it is worth obtaining the charge without the explicit need of the time dependent wave functions, as allowed by a Fourier transform based method.

As a generalization, the main topic of the manuscript is the case, when not the bare localized wave packets are investigated, but wave packets that superimpose on plane waves. The motivation for this is, in general, the local excitation of a quantum system that is described by a plane wave. (An example can be a single Bloch-state close to the minimum of a band in a semiconductor that is described using the effective mass approximation \cite{MSF18}, or a particle beam travelling in free space with low momentum uncertainty see e.g.~\cite{Echternkamp2016,Feist2015,Kozak2017,SBF17}.) In this case the relevant question is not the charge that is carried by the complete wave function (mainly because it can be infinite because of the plane wave term), but the the extra charge that is related to the presence of the wave packet. In other words, we are to determine the difference between the charges carried by the system, with and without a local excitation that creates the wave packets. A typical physical example for the excitation can be a short laser pulse that is localized to an area characterized by its focal spot size. Since the induced currents can change significantly on the femtosecond time scale, no detectors can provide time-resolved information, while the time integral of the current is usually available. The target of a laser pulse can be, e.g., a beam of charged particles or electrons in solids. For the latter case, as far as single electron approximation using a single, quadratic band is valid, the sum of contributions from all the Bloch-sates provides a charge that can be directly related to experimental results \cite{S13}.

In Sec~\ref{modelsec} we introduce a one-dimensional model in which the extra charge discussed above can be determined analytically. Then, in Sec.~\ref{Gaussec}, we focus on Gaussian wave packets with well-known time evolution, allowing the direct evaluation of the currents and charges. A comparison with our approach shows a nice agreement, which is verified also by numerical calculations. Finally, in Sec.~\ref{lasersec}, the concrete case of laser-induced currents is discussed. Conclusions are drawn in Sec.~\ref{concsec}.

\section{Analytic model}
\label{modelsec}
 Let us consider a quantum mechanical state that can be written as a sum of a  plane wave and an additional time and space dependent part. In order to simplify the calculations, we are going to use a one-dimensional description:
\begin{equation}
\Psi(x,t) = \Psi_0(x,t)+\Psi_1(x,t), \label{eq:stdiv}
\end{equation}
where $ \Psi_0(x,t) $ denotes the quantum mechanical plane wave and $ \Psi_1(x,t) $ is the additional part. In this paper we use $ x $ as a space and $ t $ as a time parameter. The actual form of $ \Psi_0(x,t) $ is given by
\begin{equation}
\Psi_0(x,t) = \exp\left[i \left( \frac{p}{\hbar} x - \omega(p) t \right) \right] = \exp\left[\frac{i}{\hbar} \left( p x - \frac{p^2}{2 m} t \right) \right], \label{eq:pw}
\end{equation}
where $ p $ denotes the momentum and $ \omega(p) $ is the dispersion relation of the free particle according to the Schrödinge equation:
\begin{equation}
\omega(p) = \dfrac{p^2}{2 m \hbar}. \label{eq:disp}
\end{equation}
Here $ m $ is the mass of the particle and $ \hbar $ is the modified Planck constant. If there is no external field, the usual probability current density in one dimension is defined as
\begin{equation}
j(x,t) = \frac{\hbar}{m} \text{Im} \left\lbrace \Psi^*(x,t) \frac{\partial \Psi(x,t)}{\partial x} \right\rbrace.
\end{equation}
If we substitute the decomposition (\ref{eq:stdiv}) to the probability current density, we can separate three different components.
\begin{eqnarray}
j(x,t) &=& \frac{\hbar}{m} \text{Im} \left\lbrace \Psi_0^*(x,t) \frac{\partial \Psi_0(x,t)}{\partial x} \right\rbrace + \frac{\hbar}{m} \text{Im} \left\lbrace \Psi_1^*(x,t) \frac{\partial \Psi_1(x,t)}{\partial x} \right\rbrace \nonumber \\
  & & \quad + \frac{\hbar}{m} \text{Im} \left\lbrace \Psi_0^*(x,t) \frac{\partial \Psi_1(x,t)}{\partial x} + \Psi_1^*(x,t) \frac{\partial \Psi_0(x,t)}{\partial x} \right\rbrace\\
  &=& j_0(x,t) + j_1(x,t) + j_c(x,t), \label{eq:current}
\end{eqnarray}
where the current density $ j_0(x,t) $ corresponding to $ \Psi_0 ,$ i.e., the plane wave, is constant in both time and space: $ j_0(x,t) = j_0 = \hbar k / m $. The other two components describe the time and space dependent dynamics.  $ j_1(x,t) $ is the current density that belongs to $ \Psi_1(x,t) $ and $  j_c(x,t) $ involves the "interference" between the two parts of the state. For the ease of identification, we call the last part as the "cross" component that is why we use the subscript $c$. Clearly, the probability current density and the charge current density are proportional. Going further along this path, the time integral of this current can be termed as a charge. From now on, we use the term "charge" in this context.

In this paper we set the initial time to be zero and we assume that the transported charge also vanishes in every spatial point at the beginning. The charge, carried by a quantum mechanical plane wave (we denote it by $ Q_0 $) , can be calculated easily at arbitrary spacetime coordinates. Our goal is to determine the charge carried by $ \Psi(x,t) $ as added to $ Q_0 $:
\begin{eqnarray}
Q_{d}(x,t\rightarrow\infty) = Q_{d}(x) &=& \int\limits_0^{\infty} j(x,t) - j_0(x,t) \; \text{d}t\\
&=& \underbrace{ \int\limits_0^{\infty} j_1(x,t) \; \text{d}t }_{ Q_1(x) } + \underbrace{ \int\limits_0^{\infty} j_c(x,t) \; \text{d}t }_{ Q_c(x) }.  \label{eq:Qddef}
\end{eqnarray}
We handle these two time integral separately.

It is known that  potential free time evolution of any given quantum state $ \psi(x) $ can be computed by the following equation
\begin{equation}
\psi(x,t) = \frac{1}{ \sqrt{2 \pi \hbar} } \int\limits_{-\infty}^{\infty} \tilde{\psi}(p) e^{i \frac{p}{\hbar} x} e^{-i \omega(p) t} \; \text{d}p, \label{eq:psifromspect}
\end{equation}
where $ \tilde{\psi}(p) $ is the initial state in momentum representation (which is the Fourier transform of $\psi(x)$) and $ \omega(p) $ is the dispersion relation defined above. It is also clear that $ \tilde{\psi}(p) $ does not change in time.  By using these forms of $ \Psi_0(x,t) $ and $ \Psi_1(x,t) $ in Eq. (\ref{eq:Qddef}), one can realize that after collecting the time dependent integrands the time integral can be computed in both parts of the $ Q_d $.

\begin{eqnarray}
Q_1(x) &= & \frac{1}{ 2 \pi m } \textup{Im} \left\lbrace \int\limits_{-\infty}^{\infty} \int\limits_{-\infty}^{\infty} \frac{i p_2}{\hbar}  \tilde{\Psi}^*_1(p_1) \tilde{\Psi}_1(p_2) e^{\frac{i}{\hbar} (p_2-p_1) x} \left[ \int\limits_{0}^{\infty} e^{-i \left[ \omega(p_2) - \omega(p_1) \right] t} \text{d}t \right] \; \text{d}p_1 \text{d}p_2\right\rbrace  \label{eq:Q1} \\
Q_c(x) &=& \frac{1}{m \sqrt{2 \pi \hbar}} \text{Im} \left\lbrace \int\limits_{-\infty}^{\infty} i p \tilde{\Psi}_1(p) e^{\frac{i}{\hbar} (p - p_0) x } \left[ \int\limits_0^{\infty} e^{- i (\omega(p) - \omega_0(p_0) ) t} \text{d}t \right] \; \text{d}p \, \right. \nonumber \\
& & + \left. \int\limits_{-\infty}^{\infty} i p_0 \tilde{\Psi}^*_1(p) e^{\frac{i}{\hbar} (p_0 - p) x } \left[ \int\limits_0^{\infty} e^{ i ( \omega(p) - \omega_0(p_0) ) t} \, \text{d}t \right] \; \text{d}p \right\rbrace \label{eq:Qc}
\end{eqnarray}
The result of the time integral is the following:
\begin{eqnarray}
\int\limits_{0}^{\infty} e^{-i \left[ \omega(p_2) - \omega(p_1) \right] t} \text{d}t &=& \pi \delta ( \omega(p_2) - \omega(p_1) ) - \frac{i}{\omega(p_2) - \omega(p_1)}.  \label{eq:timeint}
\end{eqnarray}
To keep the traceability of the derivation, in the following we compute the parts of the $ Q_d $ separately one by one. First we derive a simple analytic form of $ Q_1(x) $ which is the charge carried by $ \Psi_1(x,t) $.

After substituting Eq. (\ref{eq:timeint})  back into Eq. (\ref{eq:Q1}) and using the properties of the Dirac-delta distribution, the equation decomposes into a sum of three integrals.
\begin{eqnarray}
Q_1(x) &=& \textup{Im} \left\lbrace  \frac{1}{2} \int\limits_{-\infty}^{\infty} i \frac{ p_1 }{ |p_1| }  \tilde{\Psi}^*_1(p_1) \tilde{\Psi}_1(p_1) \; \text{d}p_1 -  \frac{1}{2} \int\limits_{-\infty}^{\infty} i \frac{ p_1 }{ |p_1| }  \tilde{\Psi}^*_1(p_1) \tilde{\Psi}_1(-p_1) e^{-\frac{i}{\hbar} 2 p_1 x} \; \text{d}p_1\right. \nonumber \\
& & \quad +  \frac{1}{\pi} \left.  \int\limits_{-\infty}^{\infty} \int\limits_{-\infty}^{\infty}  \frac{p_2}{p_2^2 - p^2_1}  \tilde{\Psi}^*_1(p_1) \tilde{\Psi}_1(p_2) e^{\frac{i}{\hbar} (p_2-p_1) x} \; \text{d}p_1 \text{d}p_2 \right\rbrace
\end{eqnarray}
At this point, we can use the partial fraction decomposition in the last double integral and observe that the imaginary parts of the integrands which have a denominator of $ p_1 + p_2,$ can be eliminated by a simple change of variables $ p1 \leftrightarrow p2.$  After taking the imaginary parts in the remaining integrals and simplify these, we can get the following expression for  $ Q_1 $:
\begin{eqnarray}
Q_1(x) &=&  \frac{1}{2} \int\limits_{-\infty}^{\infty} \frac{ p_1 }{ |p_1| }  \tilde{\Psi}^*_1(p_1) \tilde{\Psi}_1(p_1) \; \text{d}p_1 \nonumber \\
& & +  \frac{1}{2 \pi \hbar} \int\limits_{-\infty}^{\infty} \int\limits_{-\infty}^{\infty}  \frac{e^{\frac{i}{\hbar} (p_2-p_1) x} }{\frac{i}{\hbar}(p_2 - p_1)}  \tilde{\Psi}^*_1(p_1) \tilde{\Psi}_1(p_2) \; \text{d}p_1 \text{d}p_2.\label{eq:Q1_complex}
\end{eqnarray}
This equation can be simplified further if we notice that in the first integrand the $ \tilde{\Psi}^*_1(p_1) \tilde{\Psi}_1(p_1) = \tilde{\rho}(p_1) $ is the spectral density of the $ \Psi_1(p_1) $. In the second integral we can use that
\begin{equation}
\int\limits_{-\infty}^{x} e^{\frac{i}{\hbar} (p_2-p_1) s} \; \text{d}s = \frac{e^{\frac{i}{\hbar} (p_2-p_1) x}}{ \frac{i}{\hbar} (p_2-p_1) } - \underbrace{\lim\limits_{s \rightarrow - \infty}  \frac{e^{\frac{i}{\hbar} (p_2-p_1) s}}{ \frac{i}{\hbar} (p_2-p_1) } }_{ L_e(p_2,p_1) } \label{eq:expint},
\end{equation}
where $ L_e(p_2,p_1)$ is limited thus the fraction in Eq.~(\ref{eq:Q1_complex}) can be eliminated up to an additive constant. After substituting this back to the Eq.~(\ref{eq:Q1_complex}) and swap the sequence of the integrals, inverse Fourier transform of $ \Psi_1(p) $ and the conjugate of this function appears. Finally, this term can be written as an integral of the density in momentum representation up to an additional integral which contains  $L_e(p_2,p_1) \tilde{\Psi}_1(p_2) \tilde{\Psi}^*_1(p_1) $ in the integrand. This is denoted by $ I_{2,1} $.
\begin{equation}
Q_1(x) = \frac{1}{2} \int\limits_{-\infty}^{\infty} \text{sign}(p) \tilde{\rho}(p) \; \text{d}p + \int\limits_{-\infty}^x \rho(s) \;\text{d}s + I_{2,1}. \label{eq:Q1_final}
\end{equation}

\bigskip

As a next step we derive an easily computable analytic form of $ Q_c(x) $. In Eq (\ref{eq:Qc}) we can make the same steps as in the Eq (\ref{eq:Q1}). After substituting back the time integral and make the simplifications, $ Q_c(x) $ can be expressed as a sum of four different parts:
\begin{eqnarray}
Q_c(x)&=& \frac{\hbar}{\sqrt{2 \pi \hbar}} \text{Im} \left\lbrace i \pi \frac{p_0}{ |p_0| } \left[ \tilde{\Psi}_1(p_0) + \tilde{\Psi}^*_1(p_0) \right] \right.\nonumber \\
&& \qquad \qquad - i \pi \frac{p_0}{ |p_0| } \left[\tilde{\Psi}_1(-p_0) e^{-\frac{i}{\hbar} 2 p_0 x } - \tilde{\Psi}^*_1(-p_0) e^{\frac{i}{\hbar} 2 p_0 x } \right] \nonumber \\
&& \qquad \qquad + 2 \int\limits_{-\infty}^{\infty} \frac{p}{p^2 - p^2_0} \tilde{\Psi}_1(p) e^{\frac{i}{\hbar} (p - p_0) x } \; \text{d}p\nonumber \\
&& \qquad \qquad - \left. 2 \int\limits_{-\infty}^{\infty} \frac{p_0}{p^2 - p^2_0}  \tilde{\Psi}^*_1(p) e^{- \frac{i}{\hbar} (p - p_0) x } \; \text{d}p \right\rbrace.
\end{eqnarray}
The imaginary part of the second term is zero, so this can be omitted and the remaining two integrals can be combined into a single one.
\begin{equation}
Q_c(x) = \sqrt{2 \pi \hbar} \, \text{sign}(p_0) \, \text{Re} \left\lbrace  \tilde{\Psi}_1(p_0) \right\rbrace + \frac{2}{\sqrt{2 \pi \hbar}} \text{Re} \left\lbrace \int\limits_{-\infty}^{\infty} \frac{ e^{\frac{i}{\hbar} (p-p_0) x } }{\frac{i}{\hbar}(p - p_0)} \tilde{\Psi}_1(p) \; \text{d}p \right\rbrace
\end{equation}
We note that the factor $ e^{\frac{i}{\hbar} (p-p_0) x } / \left( \frac{i}{\hbar}(p - p_0) \right) $ can also be expressed by the integral plus an additive constant as we showed in Eq. (\ref{eq:expint}). Again, the sequence of the integrals can be swapped and after that one can see that the second integral is the inverse Fourier transform of $ \tilde{\Psi}_1(p) $. These steps lead to the simplest form of  the $ Q_c $ which reads
\begin{equation}
Q_c(x) = \sqrt{2 \pi \hbar} \, \text{sign}(p_0) \, \text{Re} \left\lbrace  \tilde{\Psi}_1(p_0) \right\rbrace + 2 \text{Re} \left\lbrace \int\limits_{-\infty}^{x} \Psi_1(s) e^{- \frac{i}{\hbar} p_0 s } \; \text{d}s \right\rbrace + I_1,
\end{equation}
where $  I_1 $ note the integral of $ \tilde{\Psi}_1(p) L_e(p,p_0) $.
Finally, we can summarize the results so the extra charge carried over the plane wave consists of five terms, and the real part of $ \Psi_1 $ evaluated at the momentum of the plane wave:
\begin{eqnarray}
Q_{d}(x) &=& \frac{1}{2} \int\limits_{-\infty}^{\infty} \text{sign}(p) \tilde{\rho}_1(p) \; \text{d}p + \int\limits_{-\infty}^x \rho_1(s) \;\text{d}s + I_{1,2} + \sqrt{2 \pi \hbar} \, \text{sign}(p_0) \, \text{Re} \left\lbrace  \tilde{\Psi}_1(p_0) \right\rbrace \nonumber\\ & &+ 2\, \text{Re} \left\lbrace \int\limits_{-\infty}^{x} \Psi_1(s) e^{- \frac{i}{\hbar} p_0 s } \; \text{d}s \right\rbrace + I_1.
\label{eq:Qd_final}
\end{eqnarray}
Three of these integrals and the real part of $ \Psi_1 $ are finite and spatially independent which means that if we investigate the difference between the extra charges carried additionally to the case of the  plane wave in two different space coordinates these terms disappear, and the final formula contains a sum of only two integrals:
\begin{eqnarray}
\Delta Q_d(x_2,x_1) = Q_{d}(x_2)-Q_{d}(x_1) = \frac{1}{2} \int\limits_{x_1}^{x_2} \rho_1(x) \; \text{d}x + 2 \text{Re} \left\lbrace \int\limits_{x_1}^{x_2} \Psi_1(x). e^{-\frac{i}{\hbar} p_0 x} \; \text{d}x \right\rbrace
\label{eq:Delta_Qd_final}
\end{eqnarray}
As we can see, it is only the wave function in the interval $[x_q,x_2]$ that plays a role in this expression.

\begin{figure}[ht]
\begin{center}
\includegraphics[width=8cm]{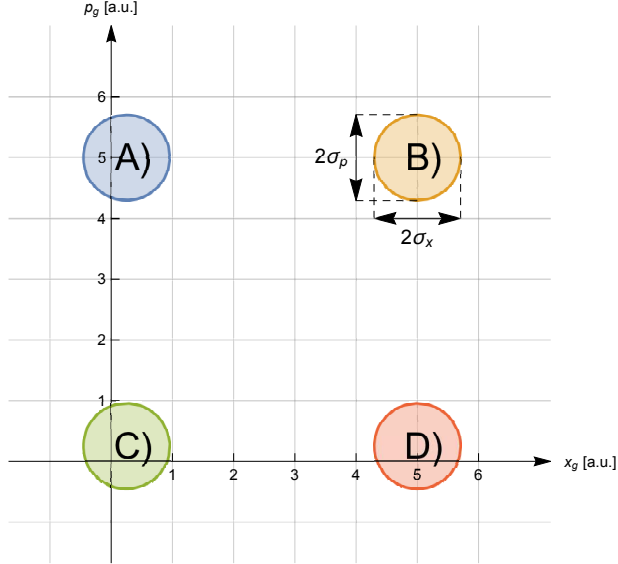}
\caption{ The investigated cases for the Gaussian wave packet example: A) $ x_G=0.25, p_G=5 $; B) $ x_G=5, p_G=5$; C) $ x_G=0.25, p_G=0.25$; D) $ x_G=5, p_G=0.25$. We use atomic units in this section, and $ \sigma_x $ and $ \sigma_p $ are the same, namely $ 1/ \sqrt{2} $.}
\label{casesfig}
\end{center}
\end{figure}

\section{An analytic example: Gaussian wave packet}
\label{Gaussec}
Let us consider the case when the additional wave packet $ \Psi_1 $ is a Gaussian cantered at $x_G.$ In momentum representation
\begin{equation}
\tilde{\Psi}_1(p) = \frac{1}{\sqrt{\sqrt{2 \pi} \sigma_p}} \exp\left( - \frac{(p-p_G)^2}{4 \sigma_p^2} - \frac{i}{\hbar} p x_G \right),
\end{equation}
where the $ \sigma_p $ is the spectral width and $ p_G $ is the mean value of the momentum distribution $ |\tilde{\Psi}_1(p)|^2 $.

The charge carried by this part in the $t\rightarrow\infty$ limit can be computed in two different ways. We can use the wave function in momentum representation as we detailed above, but we can also use the well-known time evaluation of the Gaussian wave packet and the continuity equation. Both approaches gives the same result.
\begin{equation}
Q_1 = \frac{1}{2} \text{erf} \left( \frac{x-x_0}{\sqrt{2} \sigma_x} \right) + \frac{1}{2} \text{erf}\left( \frac{p_G}{\sqrt{2} \sigma_p} \right)
\end{equation}

On the other hand, the analytical form of $ Q_c $ is the following:
\begin{align}
Q_{c}(x) = \frac{\sqrt{2 \pi \hbar}}{\sqrt{\sqrt{2 \pi} \sigma_p }} \exp\left( - \frac{(p_0-p_G)^2}{4 \sigma_p^2} \right) \bigg[ \text{Sign}(p_0) \cos\left(\frac{p_0}{\hbar} x_0 \right) \nonumber \\
 \left. + \text{Re} \left\lbrace e^{-\frac{i}{\hbar} p_0 x_0}  \text{erf}\left( \frac{x-x_0}{2 \sigma_x} + i \frac{p_0-p_G}{2\sigma_p} \right) \right\rbrace \right],
\end{align}
which can be verified by numerical computations. Using these results, we can determine $ \Delta Q_d(x_2,x_1) = Q_{d}(x_2)-Q_{d}(x_1), $ which is the charge difference between two spatial point in this Gaussian case:
\begin{eqnarray}
\Delta Q_d(x_2,x_1) &=& \frac{1}{2} \left[ \text{erf}\left( \frac{x_2-x_0}{\sqrt{2} \sigma_x} \right) - \text{erf}\left( \frac{x_1-x_0}{\sqrt{2} \sigma_x} \right) \right] + \frac{2 \sqrt{\pi} \sigma_x}{\sqrt{\sqrt{2 \pi} \sigma_x}}  \exp\left( - \frac{(p_G-p_0)^2}{ 4 \sigma_p^2 } \right) \nonumber \\
&& \text{Re} \left\lbrace e^{ - \frac{i}{\hbar} p_0 x_0} \left[ \text{erf}\left( \frac{x_2 - x_0}{2 \sigma_x} + i \frac{p_0 - p_G}{2 \sigma_p} \right) \right.\right. \nonumber\\
&& \qquad\qquad\qquad\qquad - \left. \left. \text{erf}\left( \frac{x_1 - x_0}{2 \sigma_x} +  i \frac{p_0 - p_G}{2 \sigma_p} \right) \right] \right\rbrace.
\label{eq:DQd_gauss}
\end{eqnarray}

\begin{figure}[h]
\begin{center}
\includegraphics[width=14cm]{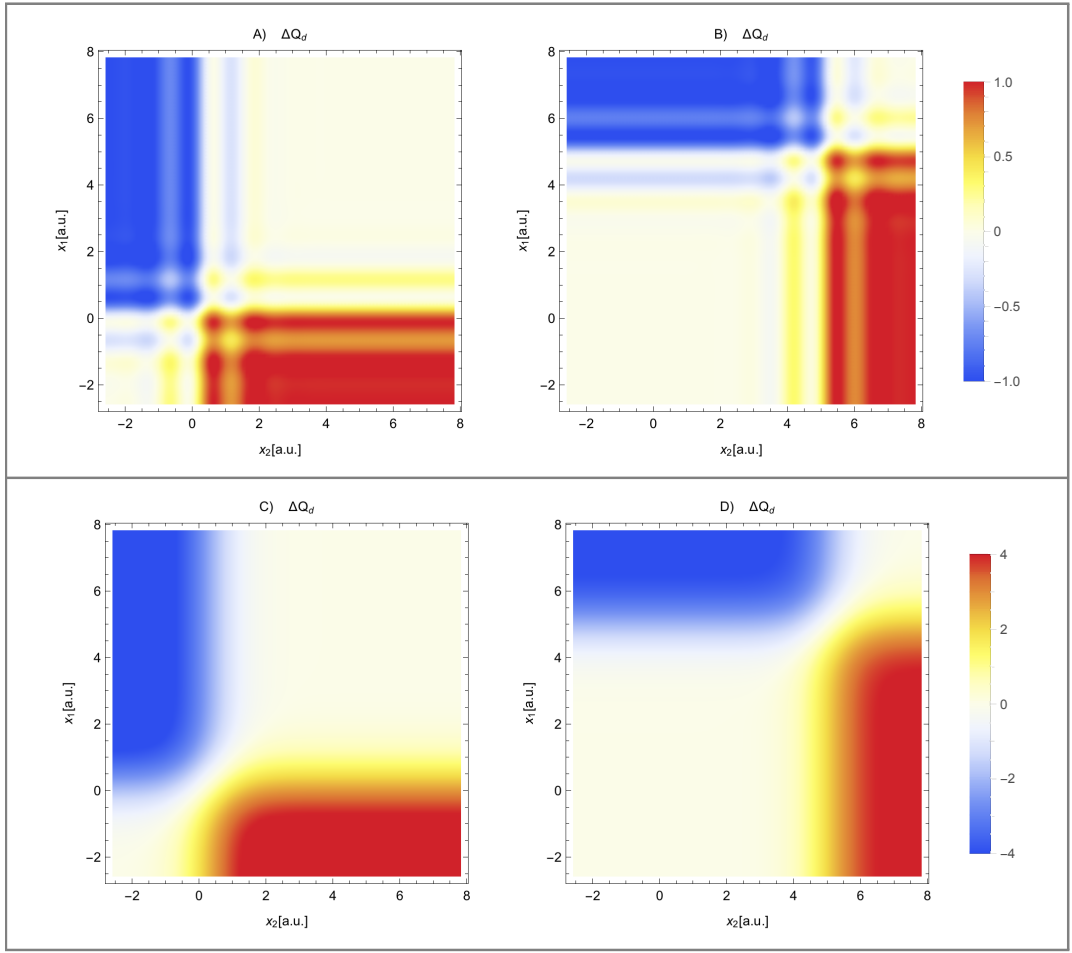}
\caption{The charge difference defined by Eq.~(\ref{eq:DQd_gauss}) between two spatial point as a function of $ x_2 $ and $ x_1 $ for the four different cases.
}
\label{xdiffig}
\end{center}
\end{figure}

We illustrate this result by considering four different cases, which are summarized in Fig.~\ref{casesfig}.
The dependence of the charge difference (\ref{eq:DQd_gauss}) of the position of the two points $x_1$ and $x_2$ is shown in Fig.~\ref{xdiffig}. As expected, $\Delta Q_d(x_2,x_1)=0$ when the interval between the points vanishes, i.e., $x_1=x_2.$ Additionally, as one can see, $\Delta Q_d(x_2,x_1)=-\Delta Q_d(x_1,x_2),$ which reflects the symmetry of the problem. When both $x_1$ and $x_2$ far from $x_G,$ the maximum of the initial Gaussian, the charge difference is practically constant, it does not depend on the positions of these points. More precisely, $\Delta Q_d(x_2,x_1)=0$ when both points are at the same side of the maximum, since in this case all the charge that flows in that direction flows through both points. On the other hand, $\Delta Q_d(x_2,x_1)$ has the maximal magnitude when the points are at the opposite side of the maximum of the initial Gaussian. When $x_1$ and $x_2$ are close to $x_G,$ interference fringes appear, but the visibility of these fringes depend on the ratio of the momenta $p_G$ and $p_0.$

\begin{figure}[h]
\begin{center}
\includegraphics[width=14cm]{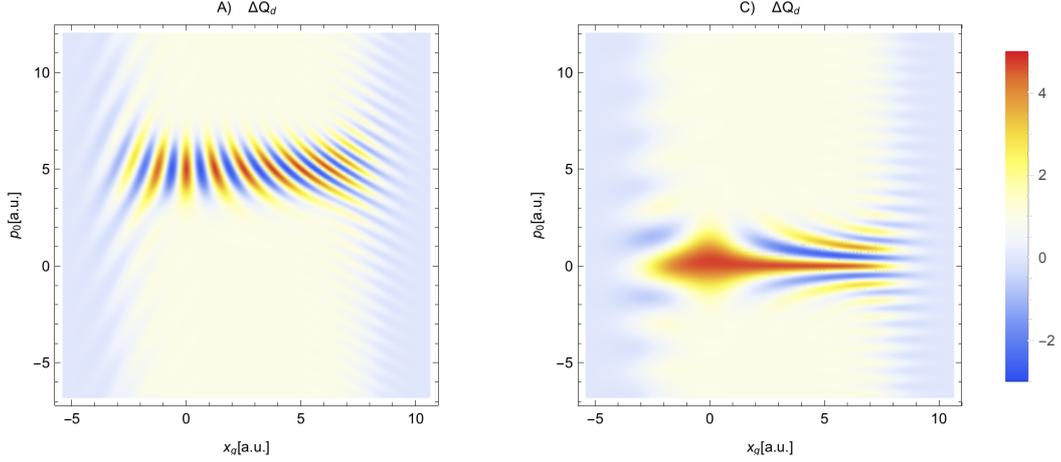}
\caption{The charge difference defined by Eq. (\ref{eq:DQd_gauss}) between two spatial points as a function of $ x_g $ and $ p_0 $ for two different cases. The first spatial parameter is fixed at $ x_2 = 7.82843$ while $ x_1 = -2.57843.$}
\label{xpdiffig}
\end{center}
\end{figure}

As analyzed by Fig.~\ref{xpdiffig}, the interference fringes become visible when $p_0,$ the momentum associated to the plane wave, is close to $p_G,$ the mean momentum of the Gaussian. In order to understood this, we have to recall that we are focusing on the long time limit. Generally, we cannot say that Fourier components with different momenta do not interfere -- clearly, their sum can produce a nontrivial pattern -- but this interference is time dependent (oscillating), and averages out when we integrate over time. On the other hand, when components with the same momentum and -- because of the same dispersion relation -- with the same frequency interfere, the resulting space domain pattern will be constant in time, and consequently survives integration over time.

\section{Numerical example: External pulse induced localized excitations}
\label{lasersec}
\begin{figure}
\begin{center}
\includegraphics[width=10cm]{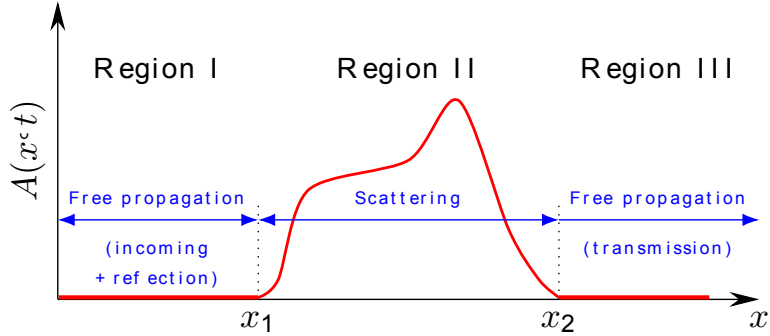}
\caption{Different spatial regions for the laser induced excitation.}
\label{regionfig}
\end{center}
\end{figure}

As a direct application of our results, we consider the case of a specific local external excitation by means of a laser pulse. For the sake of simplicity, we assume that the external electromagnetic pulse is localized both in time and space. Our previous results assume that the initial time instant is $ t=0, $ therefore the pulse starts at $ t = -\tau $ and after $ t=0 $ the external field is zero. Because of the spatial dependence, we can determine three regions in space, as shown by Fig.~\ref{regionfig}, which makes the numerical computations easier. In the second, interaction region, the Hamiltonian describing the dynamics is given by
\begin{equation}
H(x,t)=\frac{1}{2 m}(p-e{A(x,t)})^{2}, \label{Ham}
\end{equation}
where $e$ denotes the elementary charge, $m$ is the mass of the electron, $p=-i\hbar \frac{\partial}{\partial x}$ is the canonical momentum, and $A$ denote the vector potential corresponding to the excitation. $A$ is zero outside the interaction region, i.e., $H=H_0=\frac{p^2}{2m}$ in regions I and III.

We also assume that the vector potential is zero for $t<-\tau,$ therefore the initial state of the system can be a one dimensional quantum mechanical plane wave with wave number $k$ and angular frequency $\omega$, i.e.:
\begin{equation}
\Psi(x,t)=\Psi_0(x,t)=e^{i\left[k x-\omega(k) t\right]}, \ \ t<-\tau.
\end{equation}
Note that $H_0 \Psi(x,t) = E(k) \Psi(x,t),$ with $E(k)=\frac{\hbar^2 k^2}{2m^*}=\hbar \omega(k).$ For the sake of definiteness, we choose positive wave numbers $k,$ i.e., the initial plane wave propagate in the positive $x$ direction.

In our simulations the vector potential is given by
\begin{equation}
A(x,t) = A_0 \cos^2 \left( \frac{\pi}{l} (x-x_0) \right) \sin^2\left( \frac{\pi}{\tau} t \right) \cos \left( \omega_0 t \right),	
\label{vectpot}
\end{equation}
 provided $x \in [x_1,x_2]$ and $t \in [-\tau,0]$, otherwise $ A(x,t)=0$. In the simulations, we consider the central wavelength of the laser to be $ \lambda_0 = \frac{2 \pi c}{\omega_0} = 800$ nm, and note that $l=x_2-x_1$. The electric field can be calculated as the negative of the time derivative of $A(x,t)$.

When the pulse is over, at $ t=0, $ the state $\Psi(x,t=0)$  of the system can be factorized as a sum of $ \Psi_0(x,t=0) $ (plane wave) and an additional part that describes the effect of the excitation, $ \Psi_1(x,t=0) $ [recall Eq.~(\ref{eq:stdiv})]. That is,  we can use Eq.~(\ref{eq:Delta_Qd_final}) to compute charge difference as induced by the external electromagnetic pulse. Numerically, this means a simple summation and a fast Fourier transform which has low computational cost, thus it is an effective tool to investigate how the pulse parameters influence the charge difference.

\begin{figure}
\begin{center}
\includegraphics[width=10cm]{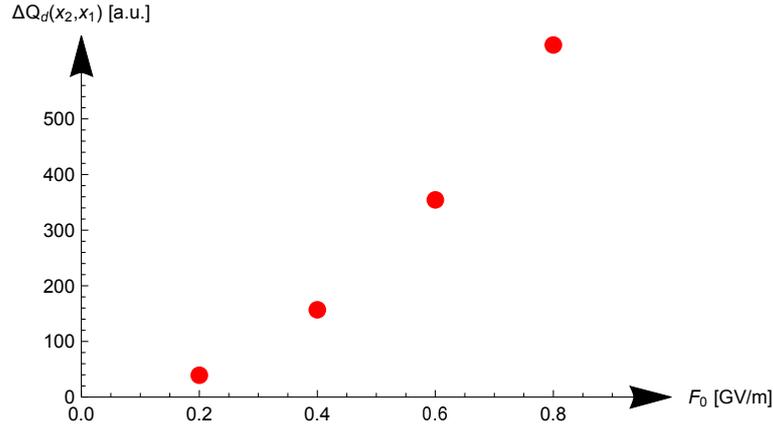}
\caption{The dependence of the charge difference between points with a distance of 800 nm as a function of the peak electric field strength $F_0.$ The central wavelength of the excitation is $c 2\pi/\omega_0=800$ nm.}
\label{Ifig}
\end{center}
\end{figure}

As an example, Fig.~\ref{Ifig} shows the dependence of the charge difference on the peak electric field strength, $F_0.$ As we can see, $\Delta Q_d(x_2,x_1)$ increases nonlinearly with $F_0.$
\section{Conclusions}
\label{concsec}
We investigated the local excitation of a physical system that can be described by a quantum mechanical plane wave. A closed formula was given for the charge difference that can be measured between two distinct spatial positions. We compared our results to the analytically solvable case of a Gaussian wave packet, and found complete agreement. First examples of the application of our approach for the description of laser induced disturbances were also shown.

\section*{Acknowledgements}
Out work was supported by the UNKP-18-3 New National Excellence Program of the Ministry of Human Capacities.

\end{document}